\documentclass[prd,superscriptaddress,showpacs,nofootinbib,amsmath,amssymb,aps,10pt]{revtex4}

\usepackage{bm}
\usepackage{amsfonts}
\usepackage{latexsym}
\usepackage[latin1]{inputenc}
\usepackage{graphicx}
\usepackage{amsmath}
\usepackage{palatino}
\usepackage{mathpazo}
\usepackage[outdir=./]{epstopdf}

\usepackage{booktabs}
\usepackage{dcolumn}

\def\jnl@style{\it}
\def\aaref@jnl#1{{\jnl@style#1}}

\def\aaref@jnl#1{{\jnl@style#1}}

\def\araa{\aaref@jnl{Ann.~Rev.~Astron.~Astrophys.}}                   
\def\aj{\aaref@jnl{AJ}}                   
\def\apj{\aaref@jnl{ApJ}}                 
\def\apjl{\aaref@jnl{ApJ}}                
\def\apjs{\aaref@jnl{ApJS}}               
\def\apss{\aaref@jnl{Ap\&SS}}             
\def\aap{\aaref@jnl{A\&A}}                
\def\aapr{\aaref@jnl{A\&A~Rev.}}          
\def\aaps{\aaref@jnl{A\&AS}}              
\def\mnras{\aaref@jnl{Mon.~Not.~Roy.~Astron.~Soc.}}             
\def\prd{\aaref@jnl{Phys.~Rev.~D}}        
\def\prc{\aaref@jnl{Phys.~Rev.~C}}  
\def\prl{\aaref@jnl{Phys.~Rev.~Lett.}}    
\def\qjras{\aaref@jnl{QJRAS}}             
\def\skytel{\aaref@jnl{S\&T}}             
\def\ssr{\aaref@jnl{Space~Sci.~Rev.}}     
\def\zap{\aaref@jnl{ZAp}}                 
\def\nat{\aaref@jnl{Nature}}              
\def\aplett{\aaref@jnl{Astrophys.~Lett.}} 
\def\apspr{\aaref@jnl{Astrophys.~Space~Phys.~Res.}} 
\def\physrep{\aaref@jnl{Phys.~Rep.}}      
\def\physscr{\aaref@jnl{Phys.~Scr}}       
\def\commat{\aaref@jnl{Comm.~Math.~Phys.}}              
\def\science{\aaref@jnl{Science}}               
\def\cqg{\aaref@jnl{Classical Quant.~Grav.}}            
\def\jpcs{\aaref@jnl{JPCS}}                                     
\def\ijmpd{\aaref@jnl{Int.~J.~Mod.~Phys.~D}}                    
\def\grg{\aaref@jnl{Gen.~Relat.~Gravit.}}               
\def\rpp{\aaref@jnl{Rep.~Prog.~Phys.}}          
\def\npa{\aaref@jnl{Nucl.~Phys.~A}}        
\def\lrr{\aaref@jnl{Living Rev.~Rel.}}                   
\def\jcap{\aaref@jnl{J.~Cosmology Astropart.~Phys.}}    
\def\rmp{\aaref@jnl{Rev.~Mod.~Phys.}}   

\allowdisplaybreaks[1]

\begin{document}

\title{Oscillation modes of rapidly rotating  neutron stars in scalar-tensor theories of gravity}
\author{Stoytcho S. Yazadjiev}
\email{yazad@phys.uni-sofia.bg}
\affiliation{Department
	of Theoretical Physics, Faculty of Physics, Sofia University, Sofia
	1164, Bulgaria}
\affiliation{Theoretical Astrophysics, Eberhard-Karls University
	of T\"ubingen, T\"ubingen 72076, Germany}

\author{Daniela D. Doneva}
\email{daniela.doneva@uni-tuebingen.de}
\affiliation{Theoretical Astrophysics, Eberhard-Karls University
	of T\"ubingen, T\"ubingen 72076, Germany}
\affiliation{INRNE - Bulgarian Academy of Sciences, 1784  Sofia, Bulgaria}

\author{Kostas D. Kokkotas}
\email{kostas.kokkotas@uni-tuebingen.de}
\affiliation{Theoretical Astrophysics, Eberhard-Karls University
	of T\"ubingen, T\"ubingen 72076, Germany}

\begin{abstract}
	We perform the first study of the oscillation frequencies of rapidly rotating neutron stars in alternative theories of gravity, focusing mainly on the fundamental $f$-modes. We concentrated on a particular class of alternative theories -- the (massive) scalar-tensor theories. The generalization to rapid rotation is important because on one hand the rapid rotation can magnify the deviations from general relativity compared to the static case and on the other hand some of the most efficient emitters of gravitational radiation, such as the binary neutron star merger remnants, are supposed to be rotating close to their  Kepler (mass-shedding) limits shortly after their formation. We have constructed several sequences of models starting from the nonrotating case and reaching up to the Kepler limit, with different values of the scalar-tensor theory coupling constant and the scalar field mass. The results show that the deviations from pure Einstein's theory can be significant especially in the case of nonzero scalar field mass. An important property of the oscillation modes of rapidly rotating stars is that they can become secularly unstable due to the emission of gravitational radiation, that is so-called Chandrasekhar-Friedman-Schutz instability. Such unstable modes are efficient emitters of gravitational radiation. Our studies show that the inclusion of nonzero scalar field would decrease the threshold value of the normalized angular momentum where this instability stars to operate, but the growth time of the instability seems to be increased compared to pure general relativity.
\end{abstract}


\maketitle

\section{Introduction}
Naturally, the perturbations of neutron stars lead to the emission of gravitational waves that can be potentially detected on Earth. That is why studying the oscillations of neutron stars is a unique way to examine their interior and to set constraints on the equation of state \cite{Andersson98a,Benhar04,Gaertig10,Doneva2013a,Bauswein2014,Takami2014}. One of the most important questions, though, is what kind of astrophysical scenarios can produce a detectable amplitude of the emitted gravitational waves. It was shown that processes such as glitches and magnetar giant flayers would hardly lead to strong enough gravitational radiation \cite{Zink2012,Lasky2011,Siegel2013,Eysden2008} which leaves little hope that the oscillation modes from nonrotating (or slowly rotating) neutron stars will be detectable \footnote{With the exception of unstable $r$-modes that will be commented below.}. This is not the case, though, if we go to rapid rotation.  The reason is that rotating neutron stars are subject to the Chandrasekhar-Friedman-Schutz (CFS) instability. This is a secular instability which develops due to the emission of gravitational waves \cite{Chandrasekhar70,Friedman78} and different modes have different thresholds for the development of the CFS instability. The rotational $r$-modes for example are generically CFS unstable, i.e. unstable for every rotational rate, while the fundamental $f$-modes are unstable only in the rapidly rotating regime. When dissipation mechanisms, such as shear or bulk viscosity, are taken into account, it turns out that the growth of modes due to CFS instability can overcome the dissipation above roughly  20\% of the Kepler  limit for $r$-modes and 80\% for the $f$-modes \cite{Doneva2013a,Andersson03}. The amplitude of the modes of course can not grow indefinitely. Instead, it would grow until a maximum saturation amplitude is reached above which nonlinear damping mechanisms, such as mode coupling and wave breaking, kick in. The saturation amplitude for the r-modes was studied in \cite{Schenk02,Arras2003,Brink2004} while the $f$-mode case was examined in \cite{Kastaun10,Pnigouras2015,Pnigouras2016}. Further studies in this direction are needed in order to give a final answer of the question of the maximum amplitude but currently it seems that at least for the $f$-modes there are sectors of the parameter space where relatively large saturation amplitude can be reached. In our paper we will focus exactly on this class of modes also because of the reason that it is expected that they will capture well the effects of general relativity (GR) modifications \cite{Sotani04,Sotani2005}.

As we commented, the rapid rotation adds very interesting and important ingredients to the problem, but on the other hand there are not many astrophysical scenarios where such rotational rates can be achieved. The primary candidates are the proto-neutron stars formed after core-collapse and especially after binary neutron star mergers. In the former case, though, the simulations show that it would be very difficult to achieve rotation close to the Kepler limit \cite{Janka2007} while in the latter case the near Kepler rotation is a generic feature of the newborn star and CFS unstable $f$-modes are a promising source of gravitational waves \cite{Lai1995,Corsi2009,Doneva2015a}. The CFS instability is a secular one that is supposed to be active for hours or days after the mergers, but shortly after the merger on dynamical timescales $f$-modes are also supposed to be excited because of the violent processes during the merger and the emitted gravitational wave signal can be used to determine the stellar parameters and to set strong constraints on the nuclear matter equation of state \cite{Bauswein2014,Takami2014}. 

Therefore, the studies of quasinormal modes of neutron stars in the static limit are a very important first step to understand the problem and make some qualitative estimations, but in order to be able to study realistic scenarios for the gravitational wave emission by CFS unstable $f$-modes, one has to go to rapid rotation. The first studies of $f$-mode oscillations of rapidly rotating neutron stars were performed in the axisymmetric case in \cite{Dimmelmeier2006} and for non-axisymmetric modes in \cite{Gaertig08,Zink10,Gaertig10,Kruger10,Gaertig11,Doneva2013a,Yoshida12}. Only the non-axisymmetric modes, though, are prone to CFS instability. Because of the complexity of the problem, most of the studies in this case were performed in the linear regime employing the Cowling approximation \cite{Gaertig08,Gaertig10,Gaertig09,Gaertig11,Kruger10,Doneva2013a}, where the spacetime degrees of freedom are neglected and only the fluid variables are evolved in time. The only exceptions are the studies in the conformal flatness approximation \cite{Dimmelmeier2006,Yoshida12} where a few exemplary sequences with polytropic equation of state where examined, and the full general relativistic nonlinear results in \cite{Zink10}. Even though, of course, it would be best to study the fully relativistic nonlinear problem, it is extremely computer and manpower demanding and it is almost impossible to use it to study in detail the parameters space. Moreover, a generalization of  such codes to alternative theories of gravity might be an extremely complicated and time consuming task. Since our goal is to perform the first study of oscillation modes of rapidly rotating neutron stars in generalized theories of gravity and to study the qualitative effects, we will also consider the perturbation equations in the Cowling approximation, that was shown to give at least good qualitative results in the nonrotating case \cite{Sotani04,Staykov2015}.

Almost all of the above outline covers the case of pure Einstein's gravity. The studies, though, show that the gravitational wave emission of oscillating neutron stars can be a powerful tool for examining the strong field regime of gravity all well \cite{Sotani04,Sotani2005,Sotani09,Silva2015,Staykov2015}. Due to the complexity of the problem all of the studies up to now were performed in the nonrotating limit. If one considers, though, values of the parameters that are in agreement with the observations, there is not much space for deviations from general relativity for most of the alternative theories. The rapid rotation provides a new window towards exploring the problem not only because the realistic astrophysical scenarios where strong gravitational radiation is emitted operate mainly in this regime, but also because it was recently shown that rapid rotation can magnify the deviations from general relativity significantly \cite{Doneva2013,Yazadjiev2015,Doneva2016}. That is why studying the neutron star oscillations in the rapidly rotating case is very important. The reason why this problem was not examined up to now is because of its complexity. For example, rapidly rotating neutron stars were constructed only in a few alternative theories of gravity \cite{Doneva2013,Yazadjiev2015,Doneva2016,Kleihaus2014,Kleihaus2016} because deriving analytically the reduced field equations and solving them numerically in very involved. Studying oscillation modes of rapidly rotating stars, especially the non-axisymmetric ones, on the other hand is also a very complicated problem even in pure GR that was addressed only recently \cite{Gaertig08,Gaertig10,Kruger10,Doneva2013a}. Our goal will be to combine these two and perform the first study of oscillating rapidly rotating neutron stars in alternative theories of gravity determining the qualitative behavior of the oscillation frequencies and the possible deviations from general relativity.

We will focus on scalar-tensor theories (STT) both with massless and with massive scalar field. These are amongst the most natural and widely used generalizations of Einstein's theory and their essence is in the inclusion of a scalar field that is a mediator of the gravitational interaction in addition to the spacetime metric. Since $f(R)$ theories are mathematically equivalent to a particular class of scalar-tensor theories with nonzero potential of the scalar field, one can expect similar qualitative behavior also for this class of theories.  We will concentrate on a particular form of the Einstein frame coupling function that leads to a theory that is perturbatively equivalent to general relativity in the weak field regime but have large differences for strong fields. Neutron stars in this class of STT were examined for the first time in \cite{Damour1993}, slowly rotating solutions were obtained in \cite{Damour1996,Sotani2012} including second order corrections in the rotational frequency \cite{Silva2015}, and rapidly rotating scalarized neutron stars where studied in \cite{Doneva2013,Doneva2016} where it was also shown for the first time that the rotation can magnify the deviation from general relativity considerably compared to the static case. We would like to also point out that the extension to massive scalar field is nontrivial and very important, because if we choose the scalar field mass in a certain limit, the current observations practically do not set any reasonable constraints on the  other parameters of the theory \cite{Popchev2015,Ramazanouglu2016,Yazadjiev2016,Doneva2016}. Thus, the deviations from pure Einstein's theory are much larger compared to the massless case.

The oscillation modes of scalar-tensor neutron stars in the nonrotating case were examined for the first time in \cite{Sotani04,Sotani2005} where the $f-$, $p-$ and the spacetime $w-$ oscillation modes were studied. The results for the $f-$ and $p-$ modes were calculated in the Cowling approximation, i.e. when the perturbations of the metric and scalar field are set to zero. It was shown that even though this approximation can deviate from the true results, it can give at least good qualitative picture. Torsional oscillations of scalarized neutron stars were examined in \cite{Silva2014}.  No oscillation modes of neutron stars in massive STT are examined up to now both in the static and rotating regimes.

The paper is organized as follows. In Section II we briefly review the background rotating neutron star solutions that we are going to perturb, in Section III the linear perturbations are discussed and in Section IV the results are presented. The paper ends with a Conclusion.

\section{Equilibrium rapidly rotating neutron star solutions in scalar-tensor theories}
Here we will review very briefly the theory behind equilibrium rotating compact star solutions in scalar-tensor theories. We refer the reader to \cite{Doneva2013,Doneva2016} for an extensive description of the problem.

All the presented equations below will be in the Einstein frame that is related to the physical Jordan frame through a conformal transformation. This frame is usually employed due to the simplicity of the field equations. All the relevant final quantities that are calculated are transformed back to the physical frame. The Einstein frame action in scalar-tensor theories has the following form
\begin{eqnarray}\label{eq:EFA}
S=\frac{1}{16\pi G} \int d^4x \sqrt{-g}\left[ R - 2
g^{\mu\nu}\partial_{\mu}\varphi \partial_{\nu}\varphi - 4 V(\varphi)
\right] + S_{\rm
	matter}(A^2(\varphi)g_{\mu\nu},\chi),
\end{eqnarray}
where $R$ is the Ricci scalar curvature with respect to the Einstein frame metric $g_{\mu\nu}$ and $V(\varphi)$ is the scalar field potential. $A(\varphi)$ is the coupling function between the matter and the scalar field\footnote{Note that such explicit coupling appears only in the Einstein frame and not in the physical Jordan frame where the weak equivalence principle is satisfied.}, and it plays the role of a conformal factor in the transformation between the Jordan frame metric ${\tilde g}_{\mu\nu}$ to the Einstein one, i.e. ${\tilde g}_{\mu\nu}= A^2(\varphi)g_{\mu\nu}$. $S_{\rm matter}$ is the action of the matter and $\chi$ denotes collectively the matter fields.

We will be working in a particular class of scalar-tensor theories that is perturbatively equivalent to general relativity but can lead to serious difference for strong field. Thus, the coupling function takes the form:
\begin{equation}\label{eq:Aphi}
\alpha(\varphi) = \frac{d\ln A(\varphi)}{d\varphi}=\beta\varphi,
\end{equation}
where $\beta$ is a constant. The reason behind this choice is that this class of scalar-tensor theories by definition fulfills the weak field observational constraints, while it can produces strong deviations from pure Einstein's theory for neutron stars. For example, for a certain range of parameters a nonlinear effect called spontaneous scalarization \cite{Damour1992,Damour1996,Doneva2013} can develop\footnote{The choice of the coupling function is by no means the only one that fulfills these requirements. As a matter of fact it is expected that other choices of $\alpha(\varphi)$ which contain odd powers of the scalar field would produce similar results.}. 

Since we will consider massive scalar-tensor theories as well, the scalar-field potential $V(\varphi)$ takes the form 
\begin{equation}
V(\varphi)=\frac{1}{2}m^2_{\varphi}\varphi^2.
\end{equation}
where $m_\varphi$ is the mass of the scalar field.

The general form of the field equations that are derived from the action (\ref{eq:EFA}) is:
\begin{eqnarray}
&&R_{\mu\nu} - \frac{1}{2}g_{\mu\nu}R= 8\pi G T_{\mu\nu} + 2\nabla_{\mu}\varphi\nabla_{\nu}\varphi - g_{\mu\nu} g^{\alpha\beta}\nabla_{\alpha}\varphi\nabla_{\beta}\varphi
- 2 V(\varphi) g_{\mu\nu}, \label{eq:FieldEq1} \\
&&\nabla_{\mu}\nabla^{\mu}\varphi = -4\pi G \alpha(\varphi) T + \frac{dV(\varphi)}{d\varphi}, \label{eq:FieldEq2}
\end{eqnarray}
where $\nabla_{\mu}$  is the covariant derivative with respect to the metric $g_{\mu\nu}$ and $T_{\mu\nu}$ is the Einstein frame energy-momentum tensor connected to the Jordan frame one ${\tilde T}_{\mu\nu}$ in the following way $T_{\mu\nu}=A^2(\varphi){\tilde T}_{\mu\nu}$. If we consider a perfect fluid the energy density, the pressure and the 4-velocity in the two  frames are connected via the coupling function $\varepsilon=A^4(\varphi){\tilde \varepsilon}$, $p=A^4(\varphi){\tilde p}$
and $u_{\mu}=A^{-1}(\varphi){\tilde u}_{\mu}$. Since the want to calculate rotating equilibrium  solutions, the spacetime, the fluid and the scalar field would be stationary and axisymmetric. Thus, the field equations can be dimensionally reduced reaching a system of several partial differential equations for the metric functions plus a second order partial differential equation for the scalar field that have to be supplemented by an equation for hydrostationary equilibrium and equation of state (EOS). Here we will not give the explicit form of the reduced field equations since they are quite lengthy and can be found in \cite{Doneva2013,Yazadjiev2015}. The general form of the metric ansatz in this case is:
\begin{eqnarray}
&&ds^2 = -e^{\gamma+\sigma} dt^2 + e^{\gamma-\sigma} r^2
\sin^2\theta (d\phi - \omega dt)^2 + e^{2\alpha}(dr^2 + r^2
d\theta^2)
\end{eqnarray}
where all metric functions depend only on the coordinates $r$ and $\theta$.  

The field equations are solved numerically using a modification of the rns code \cite{Stergioulas95} developed in \cite{Doneva2013,Yazadjiev2015,Doneva2016}. This code has proven to be accurate and reliable for constructing neutron star solutions rotating with frequencies up to the Kepler limit which makes it very useful in our case since the CFS $f$-mode instability develops especially for neutron stars rotating close to the mass-shedding limit.

In our calculations we will use a polytropic equation of state of the form
\begin{eqnarray}
&& {\tilde \varepsilon} = K \frac{{\tilde \rho}^\Gamma}{\Gamma-1} + {\tilde \rho} c^2, \nonumber \\
&& {\tilde p} = K {\tilde \rho,}^\Gamma \label{eq:PolyEOS} \\
&& \Gamma = 1+ \frac{1}{N}, \nonumber
\end{eqnarray}
where ${\tilde \rho}$ is the rest mass density in the Jordan frame, $N$ is the polytropic index and $K$ is the polytropic constant. Our calculations are performed with specific values of $N$ and $K$, namely $N=0.5$ and $K=85000$ in units $G=c=M_\odot=1$, chosen in such a way that the EOS matches well the realistic APR4 EOS. More precisely, we have adjusted these parameters in such a way that the static $M(R)$  dependence for the polytropic and realistic EOS are as close as possible. Such approach was already undertaken in other studies of neutrons star perturbations in alternative theories (see e.g. \cite{Sotani04}), since it is much easier to calculate the oscillation modes of a polytropic model especially in the rapidly rotating case.

Let us briefly comment on the observational constrains that can be imposed on the parameters of the theory. In the massless case there are very tight constraints on $\beta$ coming from the observations of neutron stars in close binary systems, namely $\beta>-4.5$ \cite{Antoniadis13,Freire2012}. Since scalarized solutions exist only for approximately $\beta<-4.35$ in the static case \cite{Harada1998} this leaves us very little freedom for variations from Einstein's theory. The rapidly rotating case produces larger deviations since it magnifies the differences from pure general relativity significantly compared to the static case and also the threshold value of $\beta$ for which scalarized solutions are present is increased. For example neutron stars with nontrivial scalar field exist for $\beta<-3.9$ in the case of rapid rotation \cite{Doneva2013}. The inclusion of nonzero scalar field mass add a highly nontrivial ingredient to the problem since the presence of mass suppresses the scalar field at length scales of the order of its Compton wavelength $\lambda_\varphi=2\pi/m_{\varphi}$. Thus for 
\begin{equation} \label{eq:bounds_mphi}
10^{-16} {\rm eV} \lesssim m_\varphi \lesssim 10^{-9}{\rm eV},
\end{equation}
corresponding to roughly $10^{-6} \lesssim m_\varphi \lesssim 10$ in our dimensionless units, the binary neutron star observations practically can not impose any constraints on coupling parameter $\beta$. More detailed information can be found in \cite{Ramazanouglu2016,Yazadjiev2016}.

In what follows, we will use the dimensionless mass of the scalar field $m_{\varphi}\to m_{\varphi} R_{0}$, where $R_{0}=1.47664 \,{\rm km}$ is one half of the solar gravitational radius.

\section{Neutron star perturbations}
The perturbation equations can be derived from the conservation law of energy momentum. In Jordan frame this law is exactly the same as in pure general relativity because of the fact that there is not a direct coupling between the matter and the scalar field (all the Jordan frame quantities below will be denoted with a tilde)
\begin{eqnarray}
{\tilde \nabla}_{\mu}{\tilde T}^{\mu}{}_{\nu} = 0,
\end{eqnarray}
where ${\tilde \nabla}_\nu$ is the covariant derivative with respect to the Jordan frame metric. As we commented above, as a first step we will adopt the Cowling approximation that is expected to give good qualitative results. In this approximation the metric and the scalar field perturbations are neglected and therefore, we are left with the same perturbation equations as in pure general relativity
\begin{equation} \label{eq:Pert_EM_Tensor}
{\tilde \nabla}_\nu (\delta {\tilde T}^{\mu\nu}) = 0\;.
\end{equation}
The perturbation of the energy momentum tensor in Cowling approximations is:
\begin{equation}
\delta {\tilde T}^{\mu\nu} = (\delta {\tilde \epsilon} + \delta {\tilde p})  {\tilde u}^\mu  {\tilde u}^\nu + ({\tilde \epsilon} + {\tilde p}) (\delta {\tilde u}^\mu {\tilde u}^\nu + {\tilde u}^\mu \delta {\tilde u}^\nu) + \delta {\tilde p}  {\tilde g}^{\mu\nu}\;.
\label{eq:pertEnergyMomentum}
\end{equation}
Here $\delta {\tilde \epsilon}$ and $\delta {\tilde p}$ are the perturbations of the energy density and the pressure related in the following way $\delta {\tilde p} = c^2_s \delta {\tilde \epsilon}$ where $ c_s$ is the speed of sound in the neutron star interior. The quantities $\delta {\tilde u}^\mu$ are the perturbations of the fluid 4-velocity that has the form ${\tilde u}^\mu = ({\tilde u}^t,0,0,{\tilde u}^\phi)$. The two nonzero components of the 4-velocity are connected via the angular frequency of the star, namely $\Omega = {\tilde u}^\phi/{\tilde u}^t$ ($\Omega$ is the same in both Jordan and Einstein frame \cite{Doneva2013}).

We will use a time evolution code in order to calculate the oscillation frequencies. A natural approach that is used in a big portion of the studies, is to evolve directly the hydrodynamical quantities  $\delta {\tilde \epsilon}$, $\delta {\tilde p}$ and $\delta {\tilde u}^\mu$. Here we will adopt a different approach that turned out to be quite useful especially in the rapidly rotating case \cite{Vavoulidis05,Vavoulidis07,Kruger10,Doneva2013a}. The perturbation variables that we will evolve in time will be the components of the energy momentum tensor. When we take into account the symmetries of the problem, the perturbation of the energy-momentum tensor can be written in the following form:
\begin{equation}
\delta T^{\mu\nu} = \left( \begin{array}{cccc} Q_1 & Q_3 & Q_4 & Q_2 \\ Q_3 & Q_6 & 0 & \Omega Q_3 \\
Q_4 & 0 & Q_6/r^2 & \Omega Q_4 \\ Q_2 & \Omega Q_3 & \Omega Q_4 & Q_5 \end{array} \right)\;,
\end{equation} 
where $Q_1,\ldots,Q_4$ are given by
\begin{eqnarray}\label{eq:Q_variables}
&& Q_1 = (\delta {\tilde \epsilon} + \delta {\tilde p}) ({\tilde u}^t)^2 + 2 ({\tilde \epsilon} + {\tilde p}) {\tilde u}^t \delta {\tilde u}^t + \delta {\tilde p} {\tilde g}^{\mu\nu}, \notag \\
&& Q_2 = (\delta {\tilde \epsilon} + \delta {\tilde p}) ({\tilde u}^t)^2 \Omega + ({\tilde \epsilon} + {\tilde p}) (\delta {\tilde u}^\phi + \Omega \delta {\tilde u}^t) {\tilde u}^t + \delta {\tilde p} {\tilde g}^{t\phi}, \\
&& Q_3 = ({\tilde \epsilon} + {\tilde p}) {\tilde u}^t \delta {\tilde u}^r,  \notag\\
&& Q_4 = ({\tilde \epsilon} + {\tilde p}) {\tilde u}^t \delta {\tilde u}^\theta\;,  \notag
\end{eqnarray}
and $Q_5$ and $Q_6$ can be expressed in terms of $Q_1$ and $Q_2$.  

The explicit form of the perturbation equations in terms of the $Q$ variables, following from equation \eqref{eq:Pert_EM_Tensor}, are quite lengthy and can be found in \cite{Kruger10,KrugerMaster}. They are solved numerically using a modification of the code developed in \cite{Kruger10,Doneva2013a}. The oscillation frequencies and the corresponding eigenfunctions can be calculated performing a Fourier transform on the computed time series. 

As we have commented in the previous section, the background neutron star solutions that will be perturbed are calculated in the Einstein frame. The perturbations equations we use, though, have a simpler form in the Jordan frame since we employ the Cowling approximation. That is why we use the transformations between the two frames given in the previous section to transform the relevant quantities between the two frame and thus the oscillations modes are calculated directly in the physical Jordan frame.
\section{Numerical results}
Here we will present the basic results for the oscillation modes of neutron stars in scalar-tensor theories. Since these are the first results in the field, as a first step we will put more emphasis on the qualitative investigation of the problem. The detailed study of the astrophysical implications, including the gravitational wave emission, will be presented in a forthcoming paper. We will focus on the case of $f$-modes and compare our results to the case of pure general relativity. We have chosen to calculate two branches with $l=m=2$ and $l=-m=2$, where $l$ is the spherical mode index and $m$ is the azimuthal mode number in the spherical harmonics\footnote{Strictly speaking we can not associate an $l$ number in the rotating case because we can not expand the oscillation modes in spherical harmonics. Instead $l$ is defined as the spherical mode index that a sequence of modes would have in the nonrotating limit.}, first because we want to investigate the so-called mode splitting in the rapidly rotating case, i.e.  breaking the degeneracy of the oscillation frequencies in the $m$ index observed in the static case, and second because especially the $l=m=2$  mode is prone to $f$-mode CFS instability and it is an important emitter of gravitational radiation.  

As we commented, in our studies we employ a polytropic equation of state that resembles the APR4 EOS in terms of the $M(R)$ dependence. This was done because in certain cases, such as small $\beta$ and strong scalar fields, the time evolution of the perturbation equations suffers from instabilities especially for rapid rotation. It is already known from the pure GR case \cite{Doneva2013a} that the mode calculation of rotating neutron star models with realistic EOS suffers in certain cases from severe problems because of the present of phase transitions inside the star. That is why, in order to be able to study well the regime of strong scalar fields, i.e. small negative $\beta$, we adopted a polytropic EOS. Since this EOS resembles the APR4 EOS, it reaches above the two solar mass barrier and has radii located in the preferred range according to the observations \cite{Demorest10,Antoniadis13,Lattimer2014,Oezel2016}. Moreover, it allows for developing of the $f$-mode CFS instability for relatively low rotational rates compared to other softer EOS \cite{Doneva2013a}.

\begin{figure}[]
	\centering
	\includegraphics[width=0.6\textwidth]{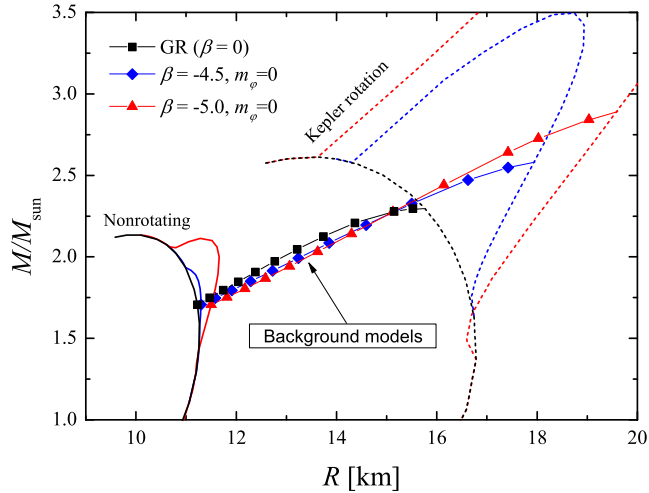}
	\caption{The mass as a function of the radius for two limiting sequences -- the static neutron stars models and the models rotating at the Kepler limit. The sequences of neutron star models that connect these two limiting cases are actually the background models that will be later perturbed in order to calculate their oscillation frequencies. The cases of pure GR, STT with $\beta=-4.5$ and with $\beta=-5$ are shown, assuming a zero scalar field mass. The case of $\beta=-6$ and nonzero $m_\varphi$ is not shown in order to have better visibility. The results are for a polytropic EOS with $N=0.5$ that has similar characteristics (such as the static $M(R)$ dependence) to the APR4 EOS.}
	\label{Fig:MR}
\end{figure}

We will study the oscillation modes of several sequences of models with constant central energy density for different values of $\beta$ and scalar field mass $m_\varphi$. Three representative examples of such sequences are shown in Fig. \ref{Fig:MR} for the cases of pure GR and massless scalar-tensor theories with $\beta=-4.5$ and $\beta=-5$. All of them stars from the nonrotating limit and reach up to the Kepler (mass-shedding) limit. The sequences of background models, that will be later perturbed in order to calculate their oscillations frequencies, are constructed in such a way that they have the same mass in the nonrotating limit. As one can see the scalar-tensor models reach in general larger masses and radii at the Kepler limit compared to the static case. Let us note that strictly speaking, in the massless case all values of $\beta$ smaller than $-4.5$ are discarded by the observations. As the results in \cite{Yazadjiev2016,Doneva2016} show, though, the case with the minimum allowed mass given by eq. \eqref{eq:bounds_mphi} for fixed $\beta$ is almost indistinguishable from the massless case with the same $\beta$, i.e. it practically reaches the maximum possible deviation. Moreover, for fixed $\beta$ the models with nonzero scalar field mass are confined between the GR models and the models with zero scalar field. Therefore, the massless case represents an upper limit of the differences from pure general relativity.
 
 \begin{figure}[]
 	\centering
 	\includegraphics[width=0.45\textwidth]{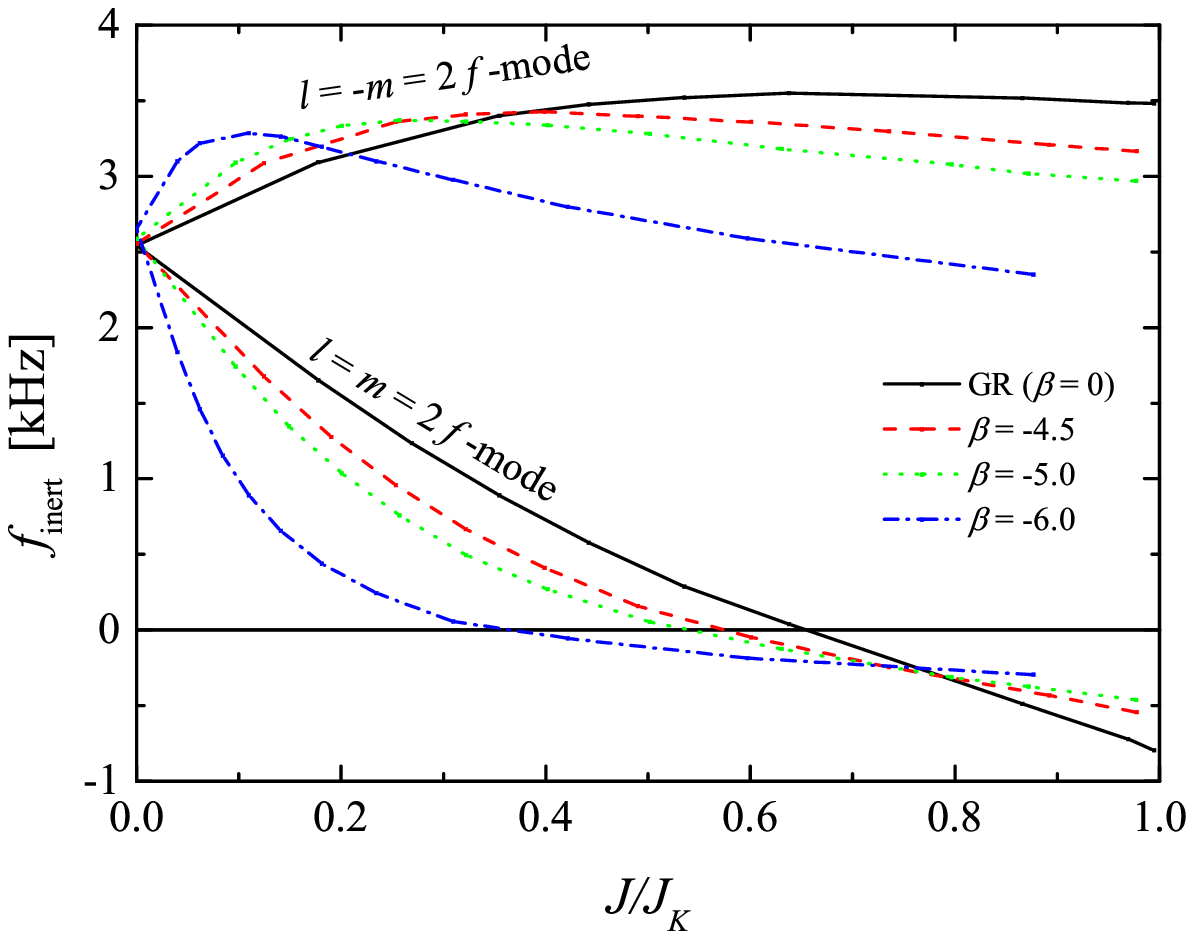}
 	\includegraphics[width=0.45\textwidth]{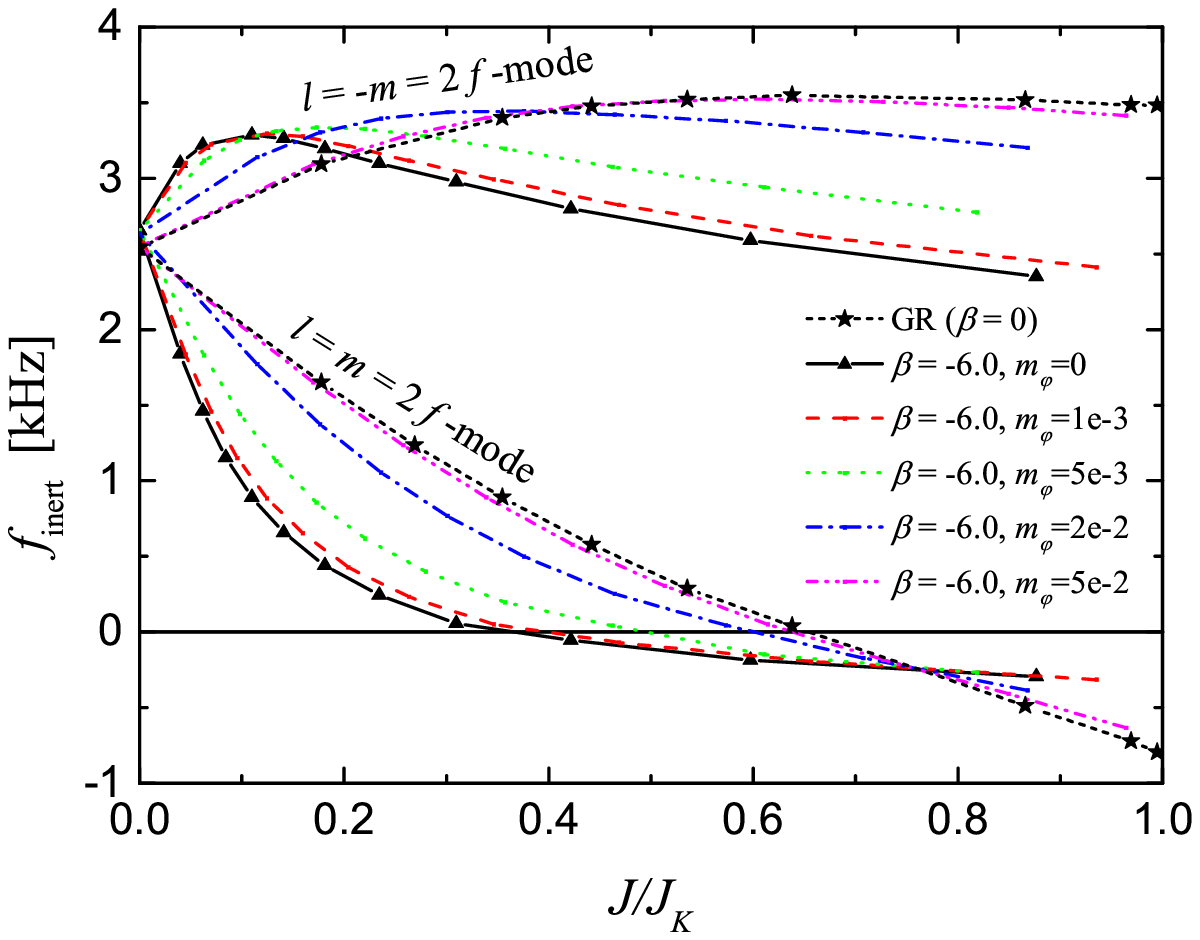}
 	\caption{The oscillation frequencies of the $l=|m|=2$ $f$-modes in inertial frame as a function of the normalized angular momentum of the star $J/J_K$ ($J_K$ is the angular momentum of the models rotating at the Kepler frequency). [Left panel] Sequences of models with different value of the coupling parameter $\beta$ in the massless scalar field case. [Right panel] Sequences of models for different values of the scalar field mass $m_\varphi$. The results are for a polytropic EOS with $N=0.5$ that has similar characteristics (such as the static $M(R)$ dependence) to the APR4 EOS.}
 	\label{Fig:QNM}
 \end{figure}

The oscillation frequencies of the modes in inertial frame, i.e. as measured from a distant observer, are plotted in Fig. \ref{Fig:QNM}. In the left panel the dependence on the coupling parameter $\beta$ in the massless scalar field case is shown while the dependence on the scalar field mass is shown in the right panel for $\beta=-6$. We have chosen to work with $\beta$ up to $-6$. As a matter of fact much smaller values of $\beta$ are allowed by the observations as long as the inequality for the scalar field mass \eqref{eq:bounds_mphi} is fulfilled that would naturally lead to considerably stronger deviations from pure GR.  On the $x$-axis the normalized moment of inertia $J/J_K$ is plotted where $J_K$ is the angular momentum of the model rotating at Kepler frequency. This choice secures us similar scales of the plots for different values of $\beta$ and $m_\varphi$ since all the sequences start from $J/J_K=0$ and span up to $J/J_K=1$. We have chosen to work with $J$ instead of the rotational frequency that was used for example in \cite{Gaertig08,Gaertig10,Doneva2013a}, because for small $\beta$ and rotation close to the Kepler limit, the rotational frequency is not a monotonous function of the angular moment (or the kinetic energy) for the sequences we have considered. Instead the rotational frequency first increases and in certain cases it starts to decrease after reaching a maximum. As a matter of fact this phenomenon also appears in pure general relativity for very large neutron star masses and rotational rates, as shown in \cite{Cook1992,Cook1994}. We should note, that some of the sequences do not reach exactly the $J/J_K=1$ limit as one can see in Fig. \ref{Fig:QNM}, because of numerical instabilities close to the Kepler limit. 

In the nonrotating case, i.e. $J/J_K=0$, the two branches of models with $m=2$ and $m=-2$ merge in a single point that represents exactly the mode degeneracy in the nonrotating limit. As the rotation is increased, the two modes split and the stable $l=-m=2$ mode increases its frequency while the frequency of the potentially unstable branch with $l=m=2$ decreases eventually crossing zero that is exactly the onset of the CFS instability. As one can see in the figure, the threshold value of $J/J_K$ for the onset of CFS instability decreases in the presence of nonzero scalar field. Let us briefly comment of the growth times of the modes. A general observation is that the more negative the frequencies are, the shorter the growth times of the instability \cite{Gaertig10,Gaertig11,Doneva2013a}. This is a consequence of the exact form of the multipole expansion of the power radiated in the form of gravitational
waves \cite{Thorne80}. Our results show that the minimum frequencies reached close to the Kepler limit are larger in STT compared to the pure GR case and therefore, it is expected that the growth times would be smaller in pure Einstein's theory that would lead to a faster development of the CFS instability .

In the right panel of Fig. \ref{Fig:QNM} the dependence on the scalar field mass is shown for $\beta=-6$. As expected, the results with different masses are within the limits of pure general relativity, corresponding roughly speaking to infinite scalar field mass, and the massless case. Thus the inclusion of a scalar field does not change qualitatively the picture, but instead allows us to consider smaller negative values of $\beta$ that can lead to very large deviations from pure general relativity.

Let us now discuss the stable $l=-m=2$ branch. A general feature of this branch in the pure general relativistic case is that the oscillation frequency first increases with the increase of the rotation and afterwards starts to slightly decrease that normally happens at large rotational rates \cite{Gaertig10,Doneva2013a}. As one can see in Fig. \ref{Fig:QNM} the behavior changes in scalar-tensor theories especially for small values of $\beta$. We still observe an increase of the oscillation frequencies and decrease after reaching certain maximum, but this maximum is much better  pronounced and happens at relatively small rotational rates. Also the minimum oscillation frequency that is reached for scalarized stars is much smaller than the pure general relativistic case.

\section{Conclusion}
In the present paper we have studied for the first time the oscillation frequencies of rapidly rotating neutron stars in alternative theories of gravity, focusing in particular on (massive) scalar-tensor theories. Several sequences of models, starting from the nonrotating limit and reaching the Kepler frequency, were constructed with different values of the coupling parameter $\beta$ and the scalar field mass $m_\varphi$. All of the sequences had the same mass in the nonrotating limit. We have employed a polytropic equation of state since the calculations are in general faster and numerically more stable in this case. The polytropic EOS is chosen to have almost the same $M(R)$ dependence in the static limit as the realistic APR4 EOS. That is why the results presented here would apply for realistic neutron star models as well.

We have focused on the nonaxisymmetric fundamental $f$-modes which are efficient emitters of gravitational radiation and they also capture well the modifications of   general relativity as the studies in the nonrotating case show \cite{Sotani04,Staykov2015}. More precisely we examined the $l=|m|=2$ modes with a special emphasis on the region where the inertial frame frequencies $f_{\rm inert}$ of the $l=m=2$ modes cross zero that is the criterion for the onset of the secular  the CFS instability. We have shown that CFS instability can develop for scalarized neutron stars for lower values of the normalized angular momentum $J/J_K$ compared to the pure general relativistic case. It is expected, though, that the growth times would be larger due to the fact that the minimum oscillations frequency reached at the Kepler limit is larger for STT compared to pure GR. The stable $l=-m=2$ branch also exhibits interesting properties especially for large values of $\beta$ -- the oscillation frequencies reach a well pronounced maximum at relatively small rotational rates and afterwards start to rapidly decrease reaching considerably smalled values compared to pure Einstein's theory.

The inclusion of mass of the scalar field does not change the picture qualitatively. For fixed $\beta$ the oscillation frequencies in the massive STT case are practically bounded between the GR limit and the massless STT case for the same value of $\beta$. Therefore, the most important contribution of the scalar field mass is that it increases significantly the range of $\beta$ that is allowed by the observations, and thus the deviations from pure GR can be much larger.

We can conclude that even though the inclusion of a scalar field increases the maximum mass of the stars especially at the Kepler limit and the onset of the CFS instability happens at smaller $J/J_K$ for the scalarized stars, the CFS instability is expected to develop on longer timescales (to have larger growth time) in STT compared to pure GR. The reason most probably lies in the fact that the CFS instability in general develops easier for more compact objects and even though the scalarized neutron stars reach larger masses, the radius increases a lot as well thus leading to effectively smaller compactness. The total mass of the star, though, is not the pure hadronic matter mass, but instead the mass contribution of the scalar field is also included. That is why such conclusions based on the compactness alone are not so straightforward. In order to compare more rigorously CFS instability in GR and STT one can consider sequences of models with constant baryon mass similar to \cite{Gaertig11,Doneva2013a,Passamonti12} and build the instability window, i.e. the range or parameters where the CFS instability overcomes the dissipative effects such as shear and bulk viscosity. Such a study is underway.

\acknowledgments{DD would like to thank the European Social Fund, the Ministry of Science, Research and the	Arts Baden-Wurttemberg and Baden W\"urttemberg Foundation for the support. The support by the Bulgarian NSF Grant DFNI T02/6, Sofia University Research Fund under Grant 80.10-30/2017, COST Action MP1304 and COST Action CA15117 is also gratefully acknowledged. }

\bibliography{references}

\end{document}